\documentclass[showpacs,prl,twocolumn,aps]{revtex4}

\usepackage{amsmath}
\usepackage{amssymb}
\usepackage{latexsym}
\usepackage{amsfonts}
\usepackage{epsfig}
\usepackage{psfrag}
\usepackage{graphicx}

\usepackage{color}

\newcommand{\bea}{\begin{eqnarray}}
\newcommand{\ea}{\end{eqnarray}}
\newcommand{\eea}{\end{eqnarray}}

\begin{document}

\title{Chiral Magnetic Spiral} 

\author{G\"ok\c ce Ba\c sar$^{1}$, Gerald~V.~Dunne$^{1}$, and Dmitri~E.~Kharzeev$^{2}$ }

\affiliation{$^1$ Department of Physics, University of Connecticut, 
Storrs CT 06269, USA
\\
$^2$Department of Physics,  Brookhaven National Laboratory, Upton, NY 11973, USA}

\begin{abstract}
We argue that the presence of a very strong magnetic field in the chirally broken phase induces inhomogeneous expectation values, of a spiral nature along the magnetic  field axis, for the currents of charge and chirality, when there is finite baryon density or an imbalance between left and right chiralities. 
This "chiral magnetic spiral" is a gapless excitation transporting the currents of (i) charge (at finite chirality), and (ii) chirality (at finite baryon density) along the direction of the magnetic field. In both cases it also induces in the transverse directions oscillating currents of charge and chirality. In heavy ion collisions, the chiral magnetic spiral possibly provides contributions both to the out-of-plane and the in-plane dynamical charge fluctuations recently observed at RHIC.
\end{abstract}



\maketitle

Recently, the STAR Collaboration at the Relativistic Heavy Ion Collider reported \cite{:2009uh, :2009txa} observation of charge-dependent azimuthal correlations, representing  evidence 
for the Chiral Magnetic Effect (CME) \cite{Kharzeev:2004ey,Kharzeev:2007tn,Kharzeev:2007jp,Fukushima:2008xe,Kharzeev:2009fn} in QCD coupled to electromagnetism. The essence of the effect is the generation of electric current along the direction of an external magnetic field in the presence of topologically nontrivial gauge field configurations creating a local imbalance between left and right chiralities. The colliding positively charged ions generate, at early times, a very strong  magnetic field, $eB \sim m_{\pi}^2$ \cite{Kharzeev:2007jp,Skokov:2009qp}, that inside the produced quark-gluon matter is directed perpendicular to the reaction plane of the collision. Topological fluctuations in the produced matter (for recent specific realizations, see \cite{Liao:2006ry,Chernodub:2009hc}) then induce the experimentally measured charge asymmetry with respect to the reaction plane that fluctuates  on an event-by-event basis.      

The CME has been studied also in lattice gauge theory, and evidence for charge separation in a magnetic field has been found both in the quenched calculation \cite{Buividovich:2009wi} and in the  calculation with dynamical light quarks \cite{Abramczyk:2009gb}. At finite baryon density, there is a closely related phenomenon of chiral separation (flow of axial current) along the direction of the magnetic field \cite{Son:2004tq,Metlitski:2005pr,Son:2009tf}. Very recently, it has been found \cite{Buividovich:2010tn} that the vacuum of the theory in the confined, chirally broken phase develops a finite electric conductivity along the direction of a sufficiently strong external magnetic field. A natural question arises about the nature of the low frequency mode capable of transporting the charge current in the vacuum. In this letter we examine the mechanism of the CME in the chirally broken phase using the method of dimensional reduction appropriate in the presence of a strong magnetic field.  
 
The two important physical effects are the lowest Landau level (LLL) projection induced by the strong magnetic field, which leads to an effective reduction of the system to $1+1$ dimensions,
as in the physics of magnetic catalysis of symmetry breaking \cite{Gusynin:1995nb,Gorbar:2009bm}, and the topological charge fluctuations of the QCD vacuum, which  induce local regions of chirality of zero modes.
\begin{figure}[htb]
\includegraphics[scale=0.3]{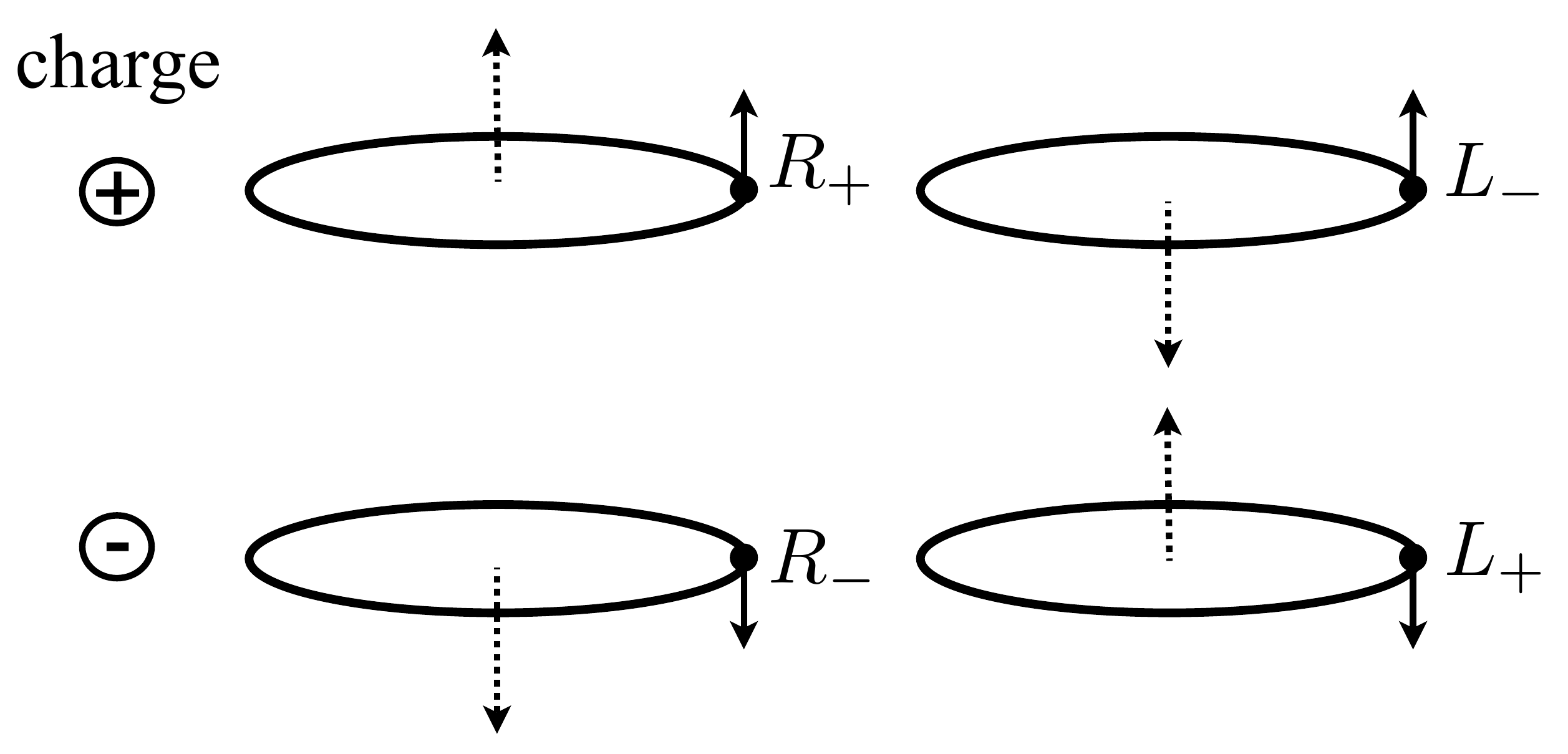}
\caption{Sketch of the effect of the strong magnetic field (directed upward in these figures) on the various spinor basis elements. The circles denote the cyclotron orbits, the spin direction is denoted by solid arrows, and the direction of momentum by dotted arrows. $R_\pm$ and $L_\pm$ label right- and left-handed chirality with $\pm$ direction of momentum along the $B$ field. The upper row, labeled on the left with a circled $+$ sign, corresponds to positive charge in the lowest Landau level projection of spins aligned along the $B$ field,  while the lower row, labeled on the left with a circled $-$ sign, corresponds to negative charge in the lowest Landau level projection of spins anti-aligned with the $B$ field. }
\label{zeeman}
\end{figure}
The chiral magnetic effect can be understood as a dimensional reduction of the chiral anomaly \cite{Adler:1969gk,Bell:1969ts,Jackiw:1983nv} to the $1+1$ dimensional chiral anomaly, expressed in terms of the $1+1$ dimensional density and current, $\chi^\dagger\chi$ and $\bar\chi \Gamma^z \chi$. Here it is crucial that the spin degeneracy of the relativistic Landau levels is absent in the LLL. Our new observation here is motivated by recent studies of chiral symmetry breaking in $1+1$ dimensional interacting fermion systems, which have highlighted the importance of a ``chiral spiral'' phase, in which the other bilinears, the scalar and pseudoscalar condensates $(\bar\chi \chi)$ and $(\bar \chi i\Gamma^5 \chi)$, become spatially inhomogeneous, forming the chiral spiral with $\Delta\equiv (\bar\chi \chi)-i(\bar \chi i\Gamma^5 \chi)=A\, e^{2i\mu z}$. This chiral spiral phase is the broken phase of the large $N_f$ chiral Gross-Neveu model \cite{Schon:2000he,Basar:2009fg}, has been identified in the 't Hooft model \cite{Bringoltz:2009ym}, and has recently been found as a 2d pion condensate in quarkyonic matter \cite{Kojo:2009ha}. In this paper we show that the chiral spiral is another facet of the chiral anomaly, and  in the presence of magnetic field it induces a chiral magnetic spiral in the longitudinal and transverse components of the axial and charge currents.

There are three important spinor bases relevant to our discussion: a chirality basis, a spin basis and a momentum direction basis,  because of the roles played by topological charge fluctuations,  Zeeman splitting, and the dimensional reduction due to lowest Landau level projection. We consider a strong (and approximately uniform) magnetic field $B$ along the $x^3$ direction, and use Dirac matrices ($j=1,2,3$)
\begin{equation}
\gamma^0=\begin{pmatrix}
0&{\bf 1}\cr
{\bf 1}&0
\end{pmatrix}\quad 
\gamma^j=\begin{pmatrix}
0&-\sigma^j\cr
\sigma^j&0
\end{pmatrix}\quad 
\gamma^5=\begin{pmatrix}
{\bf 1}&0\cr
0&-{\bf 1}
\end{pmatrix}
\end{equation}
We decompose the 4-component spinor in terms of eigenstates of the chiral projectors, $P_{R,L}=\frac{1}{2}({\bf 1}\pm \gamma^5)$, the spin projectors $P_{\uparrow,\downarrow}=\frac{1}{2}({\bf 1}\pm \Sigma^3)$, and the momentum direction projectors $P_{+,-}=\frac{1}{2}({\bf 1}\pm \gamma^0\gamma^3)$. 
The longitudinal spin operator is  
$\Sigma^3=\gamma^0\gamma^3\gamma^5 ={\rm diag}(
\sigma^3,\sigma^3)$, and the helicity operator is  
$\gamma^0\gamma^3 ={\rm diag}(
\sigma^3,-\sigma^3)$.
We can write the 4-component  spinor field as
\begin{equation}
\Psi=\begin{pmatrix}
R_+,
R_-,
L_-,
L_+
\end{pmatrix}^T
\end{equation}
Define the following 2 component spinors, eigenspinors of chirality, spin and momentum direction, respectively:
\begin{eqnarray}
&{\mathcal R}\,=\begin{pmatrix}
R_+\cr
R_-\cr
\end{pmatrix} \quad
&{\mathcal L}\,=\begin{pmatrix}
L_+\cr
L_-\cr
\end{pmatrix} 
\nonumber \\
&\phi_\uparrow=\begin{pmatrix}
R_+\cr
L_-\cr
\end{pmatrix}\quad
&\phi_\downarrow=\begin{pmatrix}
L_+\cr
R_-\cr
\end{pmatrix}
\nonumber \\
&\eta_+=\begin{pmatrix}
R_+\cr
L_+\cr
\end{pmatrix}\quad
&\eta_-=\begin{pmatrix}
L_-\cr
R_-\cr
\end{pmatrix}
\end{eqnarray}
\begin{figure}[htb]
\includegraphics[scale=0.29]{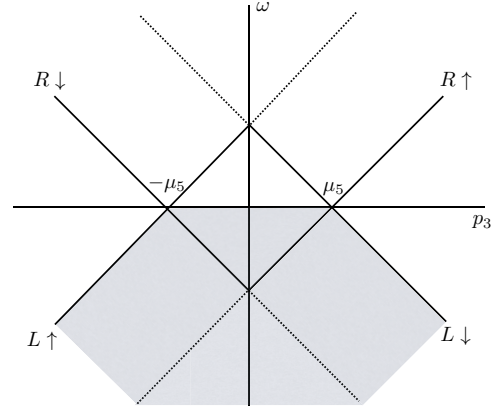}
\includegraphics[scale=0.29]{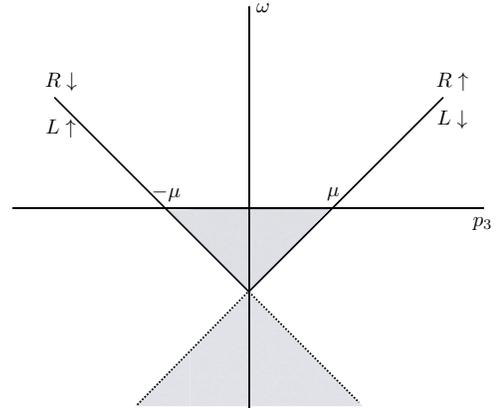}
\caption{The dispersion diagrams relevant for the case of nonzero  $\mu_5$ (upper figure) and for nonzero $\mu$ (lower figure). The associated fillings are described in the text.}
\label{fig2}
\end{figure}
Then the axial current, $J_5^\mu=\bar{\Psi}\gamma^\mu\gamma^5\Psi$, can be decomposed in terms of 2-component spinors as follows:
\begin{eqnarray}
J_5^0&=&{\mathcal R}^\dagger {\mathcal R}-{\mathcal L}^\dagger {\mathcal L}
=-i \bar\phi_\uparrow \Gamma^z \phi_\uparrow+i \bar\phi_\downarrow \Gamma^z \phi_\downarrow
\nonumber \\
J_5^1&=& \bar{\mathcal R}\,{\mathcal R}+\bar{\mathcal L}\,  {\mathcal L}
=\bar\phi_\uparrow  \phi_\downarrow+ \bar\phi_\downarrow  \phi_\uparrow
\nonumber \\
J_5^2&=&i\bar{\mathcal R}\,\Gamma^5 {\mathcal R}-i\bar{\mathcal L}\, \Gamma^5 {\mathcal L}
=-i \bar\phi_\uparrow  \phi_\downarrow+ i\bar\phi_\downarrow  \phi_\uparrow
\nonumber \\
J_5^3&=& \bar{\mathcal R}\,\Gamma^z {\mathcal R}-\bar{\mathcal L}\,\Gamma^z {\mathcal L}
=\phi_\uparrow ^\dagger \phi_\uparrow-\phi_\downarrow^\dagger \phi_\downarrow
\label{axial}
\end{eqnarray}
and the charge current, $J^\mu=\bar{\Psi}\gamma^\mu\Psi$, is decomposed:
\begin{eqnarray}
J^0&=&{\mathcal R}^\dagger {\mathcal R}+{\mathcal L}^\dagger {\mathcal L}
=\phi_\uparrow ^\dagger \phi_\uparrow+\phi_\downarrow^\dagger \phi_\downarrow
\nonumber \\
J^1&=& \bar{\mathcal R}\,{\mathcal R}-\bar{\mathcal L}\,  {\mathcal L}
=-\bar\phi_\uparrow \Gamma^5 \phi_\downarrow+ \bar\phi_\downarrow  \Gamma^5 \phi_\uparrow
\nonumber \\
J^2&=&i\bar{\mathcal R}\,\Gamma^5 {\mathcal R}+i\bar{\mathcal L}\, \Gamma^5 {\mathcal L}
=i \bar\phi_\uparrow \Gamma^5 \phi_\downarrow+ i\bar\phi_\downarrow \Gamma^5 \phi_\uparrow
\nonumber \\
J^3&=& \bar{\mathcal R}\,\Gamma^z {\mathcal R}+\bar{\mathcal L}\,\Gamma^z {\mathcal L}
=\bar{\phi_\uparrow} \Gamma^z\phi_\uparrow+\bar{\phi_\downarrow} \Gamma^z\phi_\downarrow
\label{vector}
\end{eqnarray}
Here we have defined the 2 dimensional gamma matrices as $\Gamma^0=\sigma^1, \Gamma^z=-i \sigma^2, \Gamma^5=\sigma^3$.
Note that $J^0$, $J^0_5$, $J^3$ and $J_5^3$ are expressed in terms of 2d densities and currents for ${\mathcal R}$ and ${\mathcal L}$, while $J^\perp$ and $J^\perp_5$ are expressed in terms of 2d scalar and pseudoscalar condensates for ${\mathcal R}$ and ${\mathcal L}$.

The fillings of right- and left-handed levels are shown in Figure \ref{fig2} for  the cases of nonzero chiral chemical potential $\mu_5=\mu_R=-\mu_L$, and for nonzero chemical potential $\mu=\mu_R=\mu_L$. Accounting for the appropriate branches of the excitations, the dispersion relation for  $\mu_5\neq 0$ is:
\begin{eqnarray}
\mu_5\neq 0: R\uparrow: \omega=p_3-\mu_5 \,\,\,&,&\,\,\, L\uparrow: \omega=p_3+\mu_5\nonumber\\
R\downarrow: \omega=-p_3-\mu_5  \,\,\, &,&\,\,\, L\downarrow: \omega=-p_3+\mu_5
\label{mu5disp}
\end{eqnarray}
while for $\mu\neq 0$:
\begin{eqnarray}
\mu\neq 0: R\uparrow: \omega=p_3-\mu \,\,\,&,&\,\,\, L\uparrow: \omega=-p_3-\mu\nonumber\\
R\downarrow: \omega=-p_3-\mu  \,\,\, &,&\,\,\, L\downarrow: \omega=p_3-\mu
\label{mudisp}
\end{eqnarray}
For certain pairings,   the up-moving and down-moving plane waves $e^{\pm ip_3 z}$ acquire a phase difference because of the chemical potential shift, which naturally leads to the production of sinusoidal spatial modulations in the longitudinal  direction. We show below that these modulations occur in the transverse components of the currents.

To introduce our extension of the chiral magnetic effect to the transverse components of the currents, we recall some important background for the dimensionally reduced $1+1$ dimensional fermionic theories,
 such as the chiral Gross-Neveu model. In these models, a chemical potential $\mu_f$ for a given flavor spinor $\chi$ leads to spiral behavior \cite{Schon:2000he,Basar:2009fg} of the condensate $\Delta\equiv (\bar\chi \chi)- i(\bar\chi i\Gamma^5\chi)$, which combines the scalar and pseudoscalar condensate into a single complex condensate. In these dimensionally reduced models, the density for a given flavor is simply $\rho_f=\frac{\mu_f}{\pi}$, as fixed by the $1+1$ dimensional chiral anomaly \cite{Basar:2009fg}. This follows because for a given flavor, $\chi$, a chemical potential term in the Lagrangian can be generated by the local chiral rotation $\chi\to e^{i\mu_f\Gamma^5 z}\chi$, which implies $\bar\chi\,i\partial\hskip-6pt / \chi\to \bar\chi\,i\partial\hskip-6pt / \chi-\mu_f\chi^\dagger\chi$. Consideration of the effect of such a local chiral rotation
on the {\it renormalized}  thermodynamical grand potential $\Omega^{\rm ren}[T, \mu_f]$ implies \cite{Basar:2009fg}
\begin{equation}
\Omega^{\rm ren}[T, \mu_f]=\Omega^{\rm ren}[T, \mu_f=0]-\frac{\mu_f^2}{2\pi}
\label{potential}
\end{equation}
Then the renormalized number density, $\rho_f=- \partial \Omega/\partial \mu_f$, is automatically given by $\mu_f/\pi$. Therefore, in (\ref{5lag}) with  $\rho_R=\frac{\mu_5}{\pi}$ and $\rho_L=-\frac{\mu_5}{\pi}$, combined with the Landau degeneracy factor $\frac{eB}{2\pi}$, we find from (\ref{axial}) the nonzero expectation value $\langle J_5^0\rangle = \frac{1}{2}\frac{eB}{2\pi}\frac{2 e \mu_5}{\pi}$ \footnote{In computing $\langle J_5^0\rangle$ (analogously for  $\langle J_5^3\rangle$ with $\mu\neq 0$), there is an additional factor of $\frac{1}{2}$  coming from integrating over half of the momentum space, as is appropriate for the dispersion relations associated with the relevant pairing. For a comprehensive discussion see sections II and III B in  \cite{Fukushima:2008xe}.}, while from (\ref{vector}) we have  $\langle J^0\rangle = 0$. This is another way to understand  the usual chiral magnetic effect, in a way that emphasizes the physical factorization into the product of a transverse Landau degeneracy factor and a longitudinal one-dimensional density-of-states factor. 

But this dimensionally reduced perspective also tells us something new, about the transverse components of the currents.  From the Gross-Neveu models we know that while the charge condensate $\rho_f=\frac{\mu_f}{\pi}$ is nonzero and homogeneous, the chiral condensate $\Delta$  generically has an inhomogeneous spiral behavior as soon as we introduce a chemical potential for that flavor, because the local chiral rotation generates the transformation $\Delta\to e^{2i\mu_f z} \Delta$. 
Such "chiral spiral" condensates play a key role in dimensionally reduced $1+1$ dimensional models \cite{Schon:2000he,Basar:2009fg,Bringoltz:2009ym,Kojo:2009ha}.

To understand the implications of the chiral spiral for $3+1$ dimensions,  consider first  the effect of a nonzero chiral chemical potential $\mu_5\neq 0$, corresponding to $\mu_R=\mu_5=-\mu_L$. The fillings of the fermionic single-particle levels for massless ${\mathcal R}$ and ${\mathcal L}$ particles are depicted in Figure \ref{fig2}, using the dispersion relations  in (\ref{mu5disp}).
The imbalance between right-handed and left-handed particles leads to the "chiral magnetic effect" \cite{Kharzeev:2004ey,Kharzeev:2007tn,Kharzeev:2007jp,Fukushima:2008xe,Kharzeev:2009fn}, a charge separation along the direction of the magnetic field. For example, as depicted in Figure \ref{zeeman}, in a region with a surplus of right-handed zero modes, the positive charges necessarily have spin up due to the LLL projection, and hence have positive momentum, while negatively charged right-handed particles have spin down and hence negative momentum, thereby inducing a spatial separation of charge. This charge separation is associated with a nonzero expectation value, $\langle J^3\rangle= \frac{eB}{2\pi}\frac{e \mu_5}{\pi}$, computed in \cite{Kharzeev:2004ey,Fukushima:2008xe}, reflecting the surplus of right-handed particles leading to a surplus of right-handed current along the $x^3=z$ direction. One also finds $\langle J_5^0\rangle = \frac{eB}{2\pi}\frac{e \mu_5}{\pi}$ \cite{Fukushima:2008xe,Kharzeev:2009fn}, and this is consistent with the axial anomaly, as we can view an adiabatic chiral chemical potential $\mu_5$ as producing an effective electric field $E\sim\dot{\mu}_5$, leading to the anomaly equation by the usual spectral flow picture \cite{Jackiw:1983nv,Nielsen:1983rb}. 

Our first new observation concerns the effect of a strong magnetic field and nonzero $\mu_5$ on the {\it transverse} components of the axial and charge currents. In the longitudinal direction there is no effect on the axial current, $\langle J_5^3\rangle=0$, because we assume the chirality imbalance occurs homogeneously (i.e., $\mu_5$ is almost constant along the z axis). 
On the other hand, we find nonzero expectation values for the {\it transverse} components, $\langle J_5^\perp\rangle$ and $\langle J^\perp\rangle$, of both the axial and charge currents, and moreover these have a characteristic spiral dependence on the longitudinal coordinate, set by $2\mu_5$. To see this, recall that the dispersion relations for the massless fermions are linear (\ref{mu5disp}), and the nonzero $\mu_5$ represents an imbalance in the fillings of $R$ and $L$ fermion levels, as shown in Figure \ref{fig2}. The free part of the Lagrangian with nonzero $\mu_5$ is:
\begin{eqnarray}
{\mathcal L}_{\mu_5}&=&i R_+^*\left(\partial_0+\partial_z-i\mu_5\right) R_+ 
+ i R_-^*\left(\partial_0-\partial_z-i\mu_5\right) R_- \nonumber\\
 && \hskip -.75cm
+i L_+^*\left(\partial_0+\partial_z+i\mu_5\right) L_+
+i L_-^*\left(\partial_0-\partial_z+i\mu_5\right) L_-
\label{5lag}
\end{eqnarray}

Referring to Figure \ref{fig2}, pairing occurs at the Fermi surface between ${\mathcal R}$ particles with $p_3\sim \mu_5$ and  ${\mathcal R}$ holes with $p_3\sim -\mu_5$, and between ${\mathcal L}$ particles with $p_3\sim -\mu_5$ and  ${\mathcal L}$ holes with $p_3\sim \mu_5$.
Thus the condensates $\bar{\mathcal R}\, {\mathcal R}$, $\bar{\mathcal R}\,i\Gamma^5 {\mathcal R}$, $\bar{\mathcal L}\,{\mathcal L}$ and $\bar{\mathcal L}\,i\Gamma^5{\mathcal L}$ appearing in $J^\perp$ and $J^\perp_5$ have momentum dependences displaced by $\pm \mu_5$, leading to 
\begin{eqnarray}
\langle J^1\rangle &=&C^2\cos(2\mu_5\, z-\phi_R)-D^2 \cos(2\mu_5\,z+\phi_L)
\nonumber \\
\langle J^2\rangle &=& -C^2\sin(2\mu_5\, z-\phi_R)+D^2 \sin(2\mu_5\,z+\phi_L)
\nonumber \\
\langle J_5^1\rangle &=&C^2\cos(2\mu_5\, z-\phi_R)+D^2 \cos(2\mu_5\,z+\phi_L)
\nonumber \\
\langle J_5^2\rangle &=&-C^2 \sin(2\mu_5\, z-\phi_R)-D^2 \sin(2\mu_5\,z+\phi_L)
\label{currentsmu5}
\end{eqnarray}
for some constants $C$ and $D$, and relative phases $\phi_R$ and $\phi_L$. This is consistent with the generation of chiral spiral behavior of the chiral condensates of ${\mathcal R}$ and ${\mathcal L}$ spinors.
Now consider the opposite situation where we have a particle-hole imbalance due to a real  chemical potential $\mu$. So, we take $\mu_5=0$ but $\mu\neq 0$, with  $\mu_R=\mu=\mu_L$. Then, according to \cite{Son:2004tq,Metlitski:2005pr,Son:2009tf}, there is a net flow of chirality along the direction of the magnetic field, characterized by $\langle J_5^3\rangle\neq 0$. This effect can be easily understood from (\ref{axial}) by noticing that $\langle J_5^3\rangle\neq 0$ is the difference between the spin densities. Indeed, a direct LLL computation \cite{Son:2004tq,Gorbar:2009bm} shows that   $\langle J_5^3\rangle = \frac{eB}{2\pi}\frac{e \mu}{\pi}$, in agreement with the $1+1$ dimensional charge condensate argument given above that $\rho_f=\mu_f/\pi$. In this case, as opposed to the charge separation effect when $\mu_5\neq 0$, for nonzero $\mu$ there is a separation of chirality.  On the other hand, with nonzero $\mu$, if we consider the charge current we find $\langle J^3\rangle= 0$, while $\langle J^0\rangle$ is nonzero. This is again due to assuming $\mu$ is almost constant along z, inside the collision region.  The imbalance between holes and quarks is homogeneous, resulting in vanishing current. Our second new result is that the {\it transverse} components of the axial and charge currents develop a similar spiral inhomegeneity, now characterized by $\mu$. Once again, the physical origin of the spiral dependence is the relation between the energy imbalance and the momentum imbalance through the dispersion relations for one-dimensional massless fermions. The free part of the Lagrangian with nonzero $\mu$ is:
\begin{eqnarray}
{\mathcal L}_{\mu}&=&i R_+^*\left(\partial_0+\partial_z-i\mu\right) R_+ 
+ i R_-^*\left(\partial_0-\partial_z-i\mu\right) R_- \nonumber\\
 && \hskip -.75cm
+i L_+^*\left(\partial_0+\partial_z+i\mu\right) L_+
+i L_-^*\left(\partial_0-\partial_z+i\mu\right) L_-
\label{lag}
\end{eqnarray}
Referring to Figure \ref{fig2}, pairing occurs at the Fermi surface between ${\mathcal R}$ particles with $p_3\sim \mu$ and  ${\mathcal R}$ holes with $p_3\sim -\mu$, and between ${\mathcal L}$ particles with $p_3\sim -\mu$ and  ${\mathcal L}$ holes with $p_3\sim \mu$.
Thus the condensates $\bar{\mathcal R}\, {\mathcal R}$, $\bar{\mathcal R}\,i\Gamma^5 {\mathcal R}$, $\bar{\mathcal L}\,{\mathcal L}$ and $\bar{\mathcal L}\,i\Gamma^5{\mathcal L}$ appearing in $J^\perp$ and $J^\perp_5$ have momentum dependence displaced by $\pm \mu$, leading to 
\begin{eqnarray}
\langle J^1\rangle &=&C^2\cos(2\mu z-\phi_R)-D^2 \cos(2\mu z-\phi_L)
\nonumber \\
\langle J^2 \rangle&=& -C^2\sin(2\mu z-\phi_R)-D^2 \sin(2\mu z-\phi_L)
\nonumber \\
\langle J_5^1\rangle &=&C^2\cos(2\mu z-\phi_R)+D^2 \cos(2\mu z-\phi_L)
\nonumber \\
\langle J_5^2 \rangle&=&-C^2\sin(2\mu z-\phi_R)+D^2 \sin(2\mu z-\phi_L)
\label{currentsmu}
\end{eqnarray}
As before, this spiral behavior follows  immediately from the dimensionally reduced picture, once we have an imbalance, which is here set by $\mu$.
This also implies that with both $\mu_5$ and $\mu$ being nonzero, we predict spiral condensates for the transverse components of both the axial and charge currents, with the wavenumbers being determined by both $\mu$ and $\mu_5$. 

In heavy ion collisions, the chiral magnetic spiral can induce both out-  and in-plane fluctuating charge asymmetries (the separation of out- and in-plane fluctuations has been performed recently \cite{Bzdak:2009fc} on the basis of STAR data \cite{:2009uh,:2009txa}). In the absence of topological fluctuations ($\mu_5 =0$), at finite baryon density ($\mu \neq 0$),  and in the chirally broken phase, the current of charge has only transverse  components, and the charge asymmetry will fluctuate only in-plane.  It should be kept in mind that the presence of magnetic field increases the chiral transition temperature \cite{Gusynin:1995nb}. If topological fluctuations are present in the chirally broken phase (e.g. due to the presence of meta-stable $\eta'$ domains \cite{Kharzeev:1998kz}), the  CME current can be carried by the chiral magnetic spiral. 

\smallskip

This work was supported by US DOE under grants DE-FG02-92ER40716 (GB and GD) and DE-AC02-98CH10886 (DK). We thank V. Miransky for discussions.

\end{document}